\documentclass[12pt]{article}
\usepackage[centertags]{amsmath}
\usepackage{amsfonts}
\usepackage{amssymb}
\usepackage{amsthm}
\usepackage{newlfont}
\usepackage{epsfig}
\usepackage{amscd}

\newcommand{\CC}{{\mathbb C}}
\newcommand{\beq}{\begin{equation}}
\newcommand{\eeq}{\end{equation}}
\newcommand{\ba}{\begin{array}}
\newcommand{\ea}{\end{array}}
\newcommand{\bea}{\begin{eqnarray}}
\newcommand{\eea}{\end{eqnarray}}
\begin{document}
\begin{center}
{\large \sc \bf On the relationship between  nonlinear equations integrable by the method of characteristics and equations
associated with commuting vector fields. }

\vskip 15pt

{\large  
A. I. Zenchuk$^{2,\S}$
}

\vskip 8pt



{\it  Institute of Problems of Chemecal Physics, Russian Academy of Sciences, Chernogolovka, Moscow reg., 142432}

\vskip 5pt


\vskip 5pt

{\today}

\end{center}

\begin{abstract}

 It was shown recently that
Frobenius reduction of the matrix fields reveals interesting relations among the nonlinear Partial Differential Equations (PDEs) integrable by the Inverse Spectral Transform Method ($S$-integrable PDEs), linearizable by the
 Hopf-Cole substitution ($C$-integrable PDEs) and  integrable by the method of characteristics ($Ch$-integrable PDEs). However, only two classes of $S$-integrable PDEs have been involved: soliton equations like Korteweg-de Vries, Nonlinear Shr\"odinger, Kadomtsev-Petviashvili and Davey-Stewartson equations, and $GL(N,\CC)$  Self-dual type PDEs, like Yang-Mills equation.  In this paper we consider the simple five-dimensional nonlinear PDE  from another class of $S$-integrable PDEs, namely, scalar  nonlinear  PDE  which is  commutativity condition of the pair of vector fields. We show its origin from the (1+1)-dimensional hierarchy of $Ch$-integrable PDEs after certain composition of   Frobenius type and differential reductions imposed on the matrix fields.  Matrix generalization of the above scalar nonlinear PDE will be derived as well.
\end{abstract}
%
\section{Introduction}

We continue  study of the relationship among nonlinear integrable Partial Differential Equations (PDEs) with different integrability algorithms in the spirit of  refs.\cite{Z1,ZS_2}, where  the block Frobenious   Reduction  (FR)
of the matrix fields has been introduced for this purpose.  First of all, we recall that there are several classes of completely integrable nonlinear PDEs depending on their integration algorithm. Most famous classes are following.
\begin{enumerate}
\item 
Nonlinear PDEs linearizable by some direct substitution (or $C$-integrable PDEs) \cite{Calogero1,Calogero2,Calogero3,Calogero4,Calogero5,Calogero6}. 
Mostly remarkable nonlinear PDEs of this type are PDEs linearizable by the Hopf-Cole substitution \cite{HopfCole} and by its multidimensional generalization \cite{Santini}. 
\item
Scalar equations integrable by the method of characteristics \cite{Whitham} (or $Ch$-integrable PDEs) and their matrix generalizations \cite{SZ,ZS}.
\item
Nonlinear PDEs integrable by the inverse spectral transform method  \cite{GGKM,ZMNP,AC} and by the dressing method 
 \cite{Konop,ZS1,ZS2,ZM,BM} ($S$-integrable PDEs).
We underline three subclasses of these equations:
\begin{enumerate}
\item Soliton equations in (1+1)-dimensions, such as Korteweg-de Vries (KdV) \cite{GGKM,KdV} and  Nonlinear Shr\"odinger (NLS) \cite{ZS_NLS} equations , and soliton   
(2+1)-dimensional equations, such as 
Kadomtsev-Petwiashvili (KP) \cite{KP} and  Davey-Stewartson (DS) \cite{DS} equations. Hereafter we call this subclass $S_1$-integrable PDEs.
\item
 Self-dual type PDEs having instanton solutions, like Self-dual Yang-Mills equation (SDYM) \cite{Ward,BZ}  and its multidimensional generalizations. This subclass will be called $S_2$-integrable PDEs. 
 \item
 PDEs assotiated with commutativity of vector fields \cite{Krichever,TT,DMT,KAR,GMA,MS1,MS2}. These equations may be either scalar or vector ($S_3$-integrable PDEs). 
\end{enumerate}
\end{enumerate}

It was shown in \cite{ZS_2} that $C$ and $Ch$-integrable matrix PDEs supplemented by the Frobenius reduction of the matrix fields lead to the proper class of $S$-integrable PDEs. Thus, matrix PDEs linearizable by the Hopf-Cole substitution  generate (2+1)-dimensional $S_1$-integrable PDEs; matrix equations integrable by the method of characteristics generate $S_2$-integrable PDEs or, in the particular case of constant characteristics, (1+1)-dimensional $S_1$-integrable PDEs.

 However, the technique developed in \cite{ZS_2} does not allow one to involve  $S_3$-integrable PDEs 
 into consideration.  In this paper we consider the particular example of  $S_3$-integrable PDEs which may be incorporated into the above mentioned algorithm with minor modifications. Namely, we consider the following  representative of  $S_3$-integrable PDEs \cite{MS2}:
 \begin{eqnarray}\label{intr_v}
&&
u_{z_1t_2}-u_{z_2t_1} +u_{z_2}u_{z_1x}-u_{z_1}u_{z_2x} = 0,
\end{eqnarray}
whose 
Lax pair  reads:
\begin{eqnarray}
\psi_{t_i}(\lambda;\vec x) + \lambda \psi_{z_i}(\lambda;\vec x) +u_{z_i}(\vec x) \psi_x(\lambda;\vec x) =0,\;\;i=1,2,
\end{eqnarray}  
where $\psi$ is a scalar spectral function, $\lambda$ is a spectral parameter and $\vec x=(x,z_1,z_2,t_1,t_2)$ is a set of all independent variables of nonlinear PDE.
  We show (for the restricted manifold of solutions to the eq.(\ref{intr_v})) that eq.(\ref{intr_v}) originates from the  $S_2$-integrable $GL(N^{S_2},\CC)$ SDYM  ($N^{S_2}$ is some positive integer) with independent variables $(z_1,z_2,t_1,t_2)$ after two reductions imposed on the matrix field:
 \begin{enumerate} 
 \item
 the  Differential Reduction introducing one more variable $x$  (DR($x$)) 
 \item
 the Frobenius Type Reduction (FTR) (or, in particular, just Frobenius reduction).
 \end{enumerate}
Matrix generalization of eq.(\ref{intr_v})  will   be introduced as well.  Remember that  $S_2$-integrable $GL(N^{S_2},\CC)$ SDYM originates from the (1+1)-dimensional  $Ch$-integrable hierarchy of nonlinear PDEs through the Frobenius reduction \cite{ZS_2}. All in all, the  diagram in Fig.(1)  illustrates the chain of transformations leading from the (1+1)-dimensional matrix $N\times N$ (where $N=2 N_0 n_0 M$) $Ch$-integrable hierarchy to the scalar $S_3$-integrable PDE (1).
 
\vskip 20pt
\begin{center}
\mbox{ \epsfxsize=17cm \epsffile{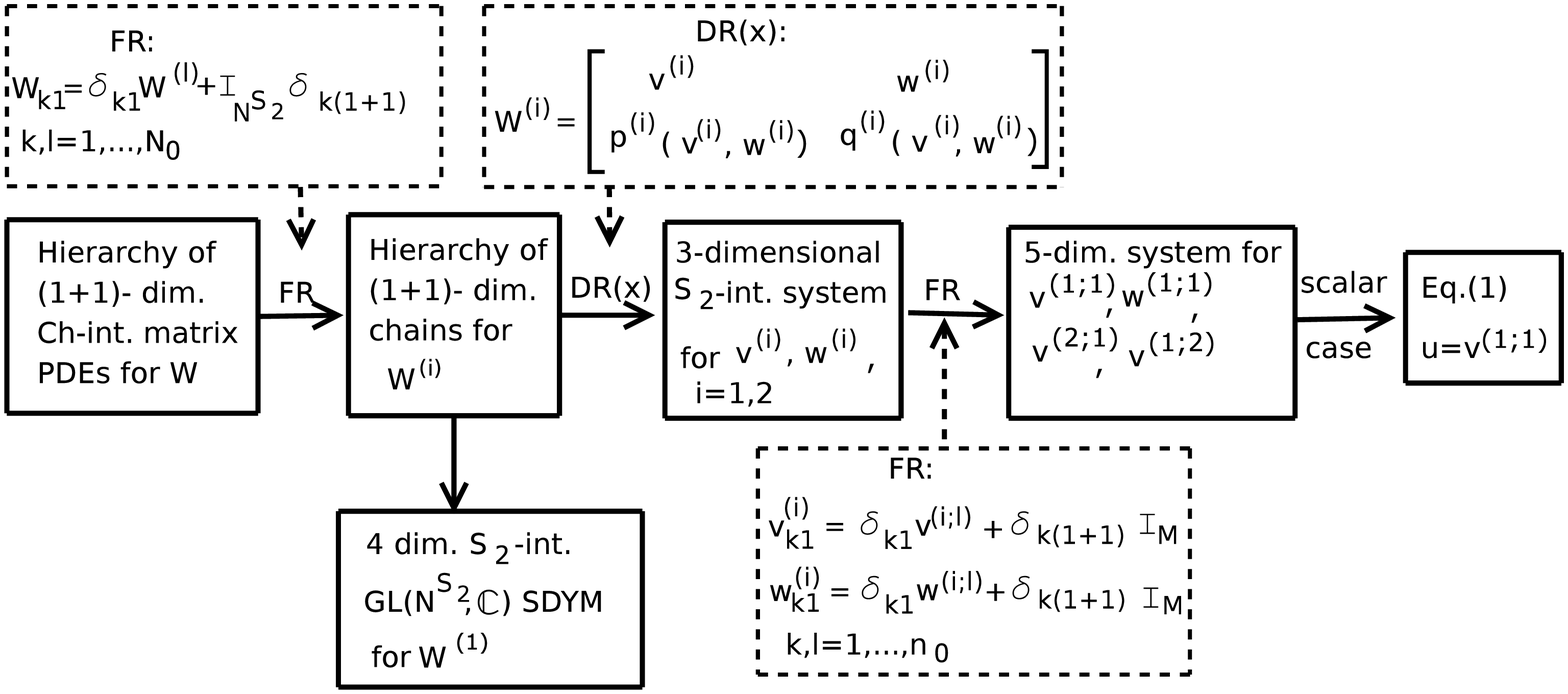}}
\end{center}
Fig.1 {\small The chain of transformations from the 
(1+1)-dimensional $Ch$-integrable hierarchy to the five-dimensional $S_3$ integrable PDE (1). Here $N=2 N_0 n_0 M$, $N^{S_2}=2 n_0 M$,  $I_{N^{S_2}}$ and $I_M$ are $N^{S_2}$- and $M$-dimensional identity matrices respectively, $W$ is $N\times N$ matrix,  $W^{(l)}$ are $2n_0 M\times 2 n_0 M$ matrices  $v^{(i)}$ and $w^{(i)}$ are $n_0 M\times n_0 M$ matrices, $v^{(i;l)}$ and $w^{(i;l)}$ are $M\times M$ matrices, $p^{(i)}$ and $q^{(i)}$ are defined by eqs.(\ref{qp_i})}
\vskip 20pt


Note that the reduction 
\begin{eqnarray}\label{red_I}
z_1=x, \;\;t_2=t,\;\; t_1=-z_2=y
\end{eqnarray}
 reduces eq.(\ref{intr_v}) into the following one
\begin{eqnarray}\label{h5_red_I}
u_{xt}+u_{yy} +u_{x}u_{xy}-u_{y}u_{xx} = 0,
\end{eqnarray}
which has been studied in \cite{P,FK,D,MS3}.

The structure of this paper is following.
In Sec.\ref{Section:general}  we recall one of the results of \cite{ZS_2}. Namely, we describe  a version of the dressing method relating  the (1+1)-dimensional $Ch$-integrable hierarchy of nonlinear  PDEs with the four-dimensional $S_2$-integrable  $GL(N^{S_2},\CC)$ SDYM. The later  admits differential reduction of the special type  introduced in Sec.\ref{Section:diff_reduction}. After that,  using FTR  we derive assotiated five-dimensional matrix system of two matrix nonlinear PDEs in 
Sec.\ref{Section:Frobenius}. Scalar case of this equation yelds the $S_3$-integrable  eq.(\ref{intr_v}).
Solution spaces to the  nonlinear PDEs derived in Secs.\ref{Section:diff_reduction} and \ref{Section:Frobenius} will be considered in Sec.\ref{Section:Solutions}. 
Conclusions are given in Sec.\ref{Section:Conclusions}.


\section{Derivation of $GL(N^{S_2},\CC)$ SDYM from the hierarchy of  matrix $Ch$-integrable nonlinear PDE.}
\label{Section:general}

In this section we describe briefly the algorithm relating the simplest (1+1)-dimensional hierarchy of  the matrix $N\times N$  $Ch$-integrable  nonlinear PDEs from one side and $GL(N^{S_2},\CC)$ SDYM from another side.  Here $N=N_0 N^{S_2}$ and $N_0$ is an arbitrary positive  integer. 
We start with the following linear equation \cite{Z1,ZS_2}
\begin{eqnarray}\label{chiW}
\chi\Lambda=W\chi
\end{eqnarray}
with $\chi$ given as  a solution to the following linear PDEs:
\begin{eqnarray}\label{chi_t}
\chi_{t_n} +\chi_{z_n}\Lambda = 0,\;\;n=1,2,\dots.
\end{eqnarray}
Here $\chi$ and $W$ are $2N_0n_0M\times 2N_0n_0M$ matrix 
functions and $\Lambda$ is a diagonal $2N_0n_0M\times 
2N_0n_0M$ matrix function.
Parameters $M$, $n_0$ and $N_0$ 
are arbitrary positive integers. Meaning of these parameters is clarified in Fig.1.
Eq.(\ref{chiW}) must be viewed as algebraic equation for $W$ with given $\chi$ and $\Lambda$:
\begin{eqnarray}\label{Wchi}
W=\chi\Lambda\chi^{-1}.
\end{eqnarray} 
Note that the diagonal matrix $\Lambda$ is not constant in general. In fact, 
compatibility condition of the eqs.(\ref{chiW}) and (\ref{chi_t})
reads
\begin{eqnarray}
\chi (\Lambda_{t_n}+\Lambda_{z_n} \Lambda) =(W_{t_n} + W_{z_n} W)\chi,\;\;n=1,2,\dots, 
\end{eqnarray}
which suggests us the following nonlinear PDEs for $\Lambda$
\begin{eqnarray}\label{nl_Lam}
\Lambda_{t_n} + \Lambda_{z_n} \Lambda =0,\;\;n=1,2,\dots
\end{eqnarray}  
and 
for $W$
\begin{eqnarray}\label{nl_W}
W_{t_n} + W_{z_n} W =0,\;\;n=1,2,\dots.
\end{eqnarray}  
Eq. (\ref{nl_Lam})
is integrable by the method of  characteristics \cite{SZ,Z1,ZS_2}.
Now both $\chi$ and $\Lambda$ are fixed, so that $W$ (solution to eq.(\ref{nl_W})) may be found using eq.(\ref{Wchi}).

Similar to  \cite{ZS_2}, in order to derive SDYM we take  $W$ in the Frobenius form:
\begin{eqnarray}\label{ww}
W&=&\left(
\begin{array}{ccccc}
W^{(1)} &W^{(2)} &\cdots& W^{(N_0-1)} & W^{(N_0)}\cr
I_{2Mn_0} &0_{2Mn_0}&\cdots & 0_{2Mn_0}& 0_{2Mn_0}\cr
0_{2Mn_0} &I_{2Mn_0}&\cdots & 0_{2Mn_0}& 0_{2Mn_0}\cr
\cdots&\cdots&\cdots&\cdots&\cdots\cr
0_{2Mn_0} &0_{2Mn_0}&\cdots & I_{2Mn_0}&0_{2Mn_0}
\end{array}
\right)
,
\end{eqnarray}
where $I_{J}$ and $0_{J}$ are $J\times J$ identity and zero matrices respectively, $W^{(i)}$ are $2 n_0M\times 2n_0M$ matrix functions.
Let matrix parameter $\Lambda$ be given in  the following block-diagonal form 
\begin{eqnarray}\label{Lambda}
&&
\Lambda={\mbox{diag}}(\Lambda^{(1)},\dots,\Lambda^{(N_0)}),
\end{eqnarray}
where 
 $ \Lambda^{(j)}$  are $2n_0M \times 2n_0M$ diagonal matrices.
Substituting (\ref{ww}) and (\ref{Lambda}) into eq.(\ref{chiW}) 
we obtain that $\chi$ must have the following block structure:
\begin{eqnarray}\label{chi}
\chi&=&\left(
\begin{array}{ccc}
\chi^{(1)} &\cdots & \chi^{(N_0)}\cr
\chi^{(1)}(\Lambda^{(1)})^{-1} &\cdots & \chi^{(N_0)}(\Lambda^{(N_0)})^{-1}\cr
\cdots&\cdots&\cdots\cr
\chi^{(1)}(\Lambda^{(1)})^{-N_0+1} &\cdots & \chi^{(N_0)}(\Lambda^{(N_0)})^{-N_0+1}
\end{array}
\right),
\end{eqnarray}
where  $\chi^{(i)}$ are $2 n_0M\times 2n_0M$ matrix functions.
In turn, $2  N_0n_0M\times 2 N_0n_0 M$ matrix equation  (\ref{chiW}) reduces to the following set of $2  n_0M\times 2  n_0M$ matrix equations:
\begin{eqnarray}\label{chi^j}
\chi^{(j)}\Lambda^{(j)} =\sum_{i=1}^{N_0} W^{(i)} \chi^{(j)}(\Lambda^{(j)})^{-i+1},\;\;j=1,\dots,N_0,
\end{eqnarray}
while eqs.(\ref{chi_t}) and (\ref{nl_Lam}) yield
\begin{eqnarray}\label{chi^j_t}
&&
\chi^{(j)}_{t_n} +\chi^{(j)}_{z_n}\Lambda^{(j)} = 0,\\
\label{nl^j_Lam}
&&
\Lambda^{(j)}_{t_n} + \Lambda^{(j)}_{z_n} \Lambda^{(j)} =0,\;\;j=1,\dots,N_0,\;\;n=1,2,\dots.
\end{eqnarray}
 Then compatibility condition of eqs. (\ref{chi^j}) and (\ref{chi^j_t}) yields 
 the following chain \cite{Z1,ZS_2}: 
\begin{eqnarray}\label{SDYM_ch}
W^{(i)}_{t_n}+W^{(1)}_{z_n} W^{(i)} +W^{(i+1)}=0, \;\;i=1,\dots, N_0,\;\;W^{(N_0+1)}=0,\;\;n=1,2,\dots,
\end{eqnarray}
which may be obtained  directly after substitution  eq.(\ref{ww}) into eq.(\ref{nl_W}).
Putting $i=1$ and eliminating $W^{(2)}$ using two equations (\ref{SDYM_ch}) with $n=1,2$ one derives 
  $GL(N^{S_2},\CC)$ SDYM, $N^{S_2}=2n_0M$:
\begin{eqnarray}\label{SDYM}
W^{(1)}_{z_nt_m}-W^{(1)}_{z_mt_n} +
W^{(1)}_{z_m}W^{(1)}_{z_n}-W^{(1)}_{z_n}W^{(1)}_{z_m}=0.
\end{eqnarray}

\section{Differential reduction of $\chi$  and assotiated system of nonlinear PDEs
}
\label{Section:diff_reduction}
Introduce one more reduction. Namely, let
matrices $\chi^{(j)}$ and $\Lambda^{(j)}$ have the following block structures:
\begin{eqnarray}\label{dif_red}
\chi^{(j)}=\left(\begin{array}{cc}
\Psi^{(2j-1)}&\Psi^{(2j)}\cr
\Psi^{(2j-1)}_x&\Psi^{(2j)}_x\cr
\end{array}\right),\;\;\Lambda^{(j)}={\mbox{diag}}(\tilde\Lambda^{(2j-1)},\tilde\Lambda^{(2j)}),\;\;
j=1,\dots,N_0,
\end{eqnarray}
where  $\Psi^{(m)}$ are $n_0M\times n_0M$ matrix functions and $\tilde\Lambda^{(m)}$ are $n_0M\times n_0M$ diagonal  matrix functions. Eqs.(\ref{chi^j_t}) and (\ref{nl^j_Lam})  yield:
\begin{eqnarray}\label{t_r1}\label{t}
&&
\Psi^{(m)}_{t_n} + \Psi^{(m)}_{z_n} \tilde \Lambda^{(m)} =0,\\\label{nl_Lam_m}
&&
\tilde \Lambda^{(m)}_{t_n} + \tilde \Lambda^{(m)}_{z_n} \tilde \Lambda^{(m)} =0,\;\;m=1,\dots,2N_0,\;\;n=1,2,\dots.
\end{eqnarray}
  $x$-dependence of $\Psi^{(m)}$ is introduced by the following second order PDE:
\begin{eqnarray}
\label{Psi_xx}\label{g_r1_3_2}\label{g_r2}\label{x}
{\cal{E}}^{(0)}&:=&\Psi^{(m)}_{xx}= a \Psi^{(m)} \tilde \Lambda^{(m)}+\nu \Psi^{(m)}_x + \mu\Psi^{(m)},\;\;m=1,\dots,2N_0,
\end{eqnarray}
where $a$, $\nu$ and $\mu$ are  $n_0M\times n_0M$ diagonal matrix parameters. These parameters must be constant and $\tilde \Lambda^{(m)}$ must be independent on $x$. In fact, compatibility condition of (\ref{t}) and (\ref{x}) yields, along with (\ref{nl_Lam_m}), the following conditions:
\begin{eqnarray}\label{Lam_x}
\tilde \Lambda^{(m)}_x=0,\;\;m=1,\dots ,2N_0,\;\;a, \;\;\nu,\;\;\mu \;\;{\mbox{are constant diagonal matrices}},\;\;m=1,\dots,N_0.
\end{eqnarray}

The block structures of $\chi^{(j)}$ and $\Lambda^{(j)}$ (\ref{dif_red}) suggest us the  relevant  block structures of $W^{(j)}$:
\begin{eqnarray}\label{ww_i}
&&
W^{(j)}=\left(\begin{array}{cc}
w^{(j)} & v^{(j)}\cr
p^{(j)} & q^{(j)}
\end{array}\right),
\end{eqnarray}
where $w^{(j)}$, $v^{(j)}$, $q^{(j)}$ and $p^{(j)}$ are $ n_0M\times n_0 M$ matrix functions.
Now set of $2 n_0  M\times 2 n_0 M$   eqs.(\ref{chi^j}) may be written as two $ n_0M \times 2N_0 n_0 M $ equations:
\begin{eqnarray}\label{g_r1}\label{b1}
&&
{\cal{E}}^{(1)}:=\Psi \Lambda =\sum_{i=1}^{N_0}\Big(
v^{(i)} \Psi_x +w^{(i)} \Psi\Big)\Lambda^{-i+1}, \\\label{b2}
&&
{\cal{E}}^{(2)}:=\Psi_{x} \Lambda =\sum_{i=1}^{N_0}\Big(
q^{(i)} \Psi_x +p^{(i)} \Psi\Big)\Lambda^{-i+1},
\end{eqnarray}
where
\begin{eqnarray}\label{Psi^}
&&
\Psi=(\Psi^{(1)},\dots,\Psi^{(2 N_0)}).
\end{eqnarray}
Compatibility condition of  eqs.(\ref{b1}) and (\ref{b2}),
\begin{eqnarray}
{{\cal{E}}^{(1)}_x=\cal{E}}^{(2)},
\end{eqnarray}
yields the expressions for $p^{(j)}$ and $q^{(j)}$ in terms of $v^{(j)}$ and $w^{(j)}$:
\begin{eqnarray}\label{qp_i}
&&
p^{(j)}=
w^{(j)}_x+v^{(j)}\mu + v^{(j+1)}  a+v^{(1)} a w^{(j)} ,\;\;\;q^{(j)}= v^{(j)}_x +v^{(j)} \nu + w^{(j)} + v^{(1)}  a v^{(j)}.
\end{eqnarray}
 Thus only two blocks of  $W^{(j)}$ are independent, i.e. $w^{(j)}$ and $v^{(j)}$.
 
Matrix equation (\ref{g_r1}) may be considered  as the uniquely solvable  system of scalar linear algebraic equations 
for the functions $v^{(i)}$ and $w^{(i)}$, $i=1,\dots,N_0$, while eq.(\ref{b2}) is the consequence of  eq.(\ref{g_r1}).
In order to derive the nonlinear PDEs for $v^{(i)}$ and $w^{(i)}$, we turn to the 
compatibility condition of the eqs.(\ref{t}) and (\ref{g_r1}):
\begin{eqnarray}
&&
{\cal{E}}^{(1)}_{t_n}+{\cal{E}}^{(1)}_{z_n}\Lambda\;\;\Rightarrow\;\;
\sum_{i=1}^{N_0}E^{(1i)} \Psi_x\Lambda^{-i+1}
+\sum_{i=1}^{N_0}E^{(0i)} \Psi\Lambda^{-i+1}
  =0,\;\;\;n=1,2,\dots,
\end{eqnarray}
which generates  the following chains of  nonlinear PDEs  for $v^{(i)}$ and $w^{(i)}$, $i=1,\dots,N_0$:
\begin{eqnarray}\label{nl1}
E^{(1i)}_n&:=&v^{(i)}_{t_n} +v^{(1)}_{z_n} (w^{(i)} +v^{(i)}_x + v^{(i)}\nu + v^{(1)}a v^{(i)})+w^{(1)}_{z_n} v^{(i)}+v^{(i+1)}_{z_n}=0,\\\nonumber
&&
v^{(N_0+1)}=0.\\\label{nl2}
E^{(0i)}_n&:=&w^{(i)}_{t_n}  + v^{(1)}_{z_n} (w^{(i)}_x+v^{(i)} \mu+v^{(1)}a w^{(i)}+v^{(i+1)}a)+w^{(1)}_{z_n} w^{(i)}+w^{(i+1)}_{z_n}=0,\\\nonumber
 &&
w^{(N_0+1)}=0,\;\;n=1,2,\dots.\\\nonumber
\end{eqnarray}
In addition, we must take into account the compatibility condition of the eqs.(\ref{g_r2}) and (\ref{g_r1})
\begin{eqnarray}\label{constr1}
{\cal{E}}^{(1)}_{xx}&=&{\cal{E}}^{(0)}\Lambda\;\;\Rightarrow\;\;
\sum_{i=0}^{N_0}(\tilde E^{(1i)}\Psi_{x}+\tilde
E^{(0i)}\Psi)\Lambda^{-i+1} =0
,
\end{eqnarray}
which gives us the non-evolutionary part of the  system of nonlinear chains, $i=1,\dots,N_0$:
\begin{eqnarray}\label{nl3}
\tilde E^{(1i)}&:=&v^{(i)}_{xx} +[w^{(i)},\nu] + 2 w^{(i)}_x + 2 v^{(i)}_x \nu-\nu v^{(i)}_x +[v^{(i)},\mu] + [v^{(i)},\nu]\nu +[v^{(1)},\nu]a v^{(i)} +\\\nonumber
&&
2 v^{(1)}_x a v^{(i)} +
[w^{(1)},a] v^{(i)}  +
[v^{(1)},a](
w^{(i)} +v^{(i)}_x + v^{(i)}\nu + v^{(1)}a v^{(i)})+[v^{(i+1)},a]=0,\\\label{nl4}
\tilde E^{(0i)}&:=&
w^{(i)}_{xx} + [w^{(i)},\mu] + [w^{(1)},a] w^{(i)} + 2 v^{(i)}_x\mu + [v^{(i)},\nu]\mu + [v^{(1)},\nu] a w^{(i)} + \\\nonumber
&&
2 v^{(1)}_xa w^{(i)} -\nu w^{(i)}_x +[v^{(1)},a](w^{(i)}_x + v^{(i)}\mu + v^{(1)} a w^{(i)} +v^{(i+1)} a) + [w^{(i+1)}, a] +\\\nonumber
&& 
2 v^{(i+1)}_x a +[v^{(i+1)},\nu] a =0
\end{eqnarray}
Of course, the system (\ref{nl1},\ref{nl2},\ref{nl3},\ref{nl4}) may be derived directly from the system (\ref{SDYM_ch}) using reduction (\ref{ww_i}). However, the derivation of this system as the compatibility condition of the linear system (\ref{t},\ref{g_r2},\ref{g_r1}) is more illustrative.

The complete system of nonlinear PDEs for $v^{(i)}$ and $w^{(i)}$, $i=1,2$,
is represented by the eqs.(\ref{nl1},\ref{nl2}) with  fixed
 $n$ (say $n=1$) and $i=1$ and by the eqs.(\ref{nl3},\ref{nl4}) with $i=1$:
\begin{eqnarray}\label{nl1_1}
E^{(11)}_1&:=&
v^{(1)}_{t_1} +v^{(1)}_{z_1} (w^{(1)} +v^{(1)}_x + v^{(1)}\nu + v^{(1)} a v^{(1)})+w^{(1)}_{z_1} v^{(1)}+v^{(2)}_{z_1}=0,\\\label{nl2_1}
E^{(01)}_1&:=&
w^{(1)}_{t_1}  + v^{(1)}_{z_1} (w^{(1)}_x+v^{(1)} \mu+v^{(1)} a w^{(1)}+v^{(2)} a)+w^{(1)}_{z_1} w^{(1)}+w^{(2)}_{z_1}=0,
\end{eqnarray}
\begin{eqnarray}\label{nl3_1}
\tilde E^{(11)}&:=&
v^{(1)}_{xx} +[w^{(1)},\nu] + 2 w^{(1)}_x + 2 v^{(1)}_x \nu-\nu v^{(1)}_x +[v^{(1)},\mu] +\\\nonumber
&&
 [v^{(1)},\nu](\nu + a v^{(1)}) +2 v^{(1)}_x  a v^{(1)} +\\\nonumber
&&
[w^{(1)}, a] v^{(1)}  +
[ v^{(1)},a](
w^{(1)} +v^{(1)}_x + v^{(1)}\nu + v^{(1)} a v^{(1)})+[v^{(2)}, a]=0,\\\label{nl4_1}
\tilde E^{(01)}&:=&
w^{(1)}_{xx} + [w^{(1)},\mu] + [w^{(1)}, a] w^{(1)} + 2 v^{(1)}_x\mu + [v^{(1)},\nu]\mu + [v^{(1)},\nu]  a w^{(1)} + \\\nonumber
&&
2 v^{(1)}_x a w^{(1)} -\nu w^{(1)}_x +[ v^{(1)},a](w^{(1)}_x + v^{(1)}\mu + v^{(1)} a w^{(1)} +v^{(2)} a) + [w^{(2)}, a] +\\\nonumber
&& 
2 v^{(2)}_x a +[v^{(2)},\nu] a =0
\end{eqnarray}
The scalar version ($n_0M=1$) of the system (\ref{nl1_1}-\ref{nl4_1}) reads:
\begin{eqnarray}\label{nl1_1s}
E_1&=&v^{(1)}_{t_1} +v^{(1)}_{z_1} (w^{(1)} +v^{(1)}_x +\nu v^{(1)} +  a v^{(1)} v^{(1)})+w^{(1)}_{z_1} v^{(1)}+v^{(2)}_{z_1}=0,\\\label{nl2_1s}
&&
w^{(1)}_{t_1}  + v^{(1)}_{z_1} (w^{(1)}_x+v^{(1)} \mu+ a v^{(1)} w^{(1)}+ a  v^{(2)})+w^{(1)}_{z_1} w^{(1)}+w^{(2)}_{z_1}=0,\\\label{nl3_1s}
&&
v^{(1)}_{xx}  + 2 w^{(1)}_x +  v^{(1)}_x \nu  +2  a v^{(1)}_x v^{(1)} 
=0,\\\label{nl4_1s}
&&
w^{(1)}_{xx} + 2 v^{(1)}_x\mu  + 
2 a  v^{(1)}_x w^{(1)} -\nu w^{(1)}_x  +
2 a v^{(2)}_x  =0
\end{eqnarray}

\section{Frobenius type reduction and assotiated higher dimensional systems of nonlinear PDEs}
\label{Section:Case1_F} 
\label{Section:Frobenius}
It is remarkable that the chains of nonlinear PDEs (\ref{nl1},\ref{nl2},\ref{nl3},\ref{nl4}) admit  the following FTR:
\begin{eqnarray}\label{G_F}
 &&
 v^{(i)}=\{v^{(i;kl)},\;\;k,l=1,\dots,n_0\},\;\;
  w^{(i)}=\{w^{(i;kl)},\;\;k,l=1,\dots,n_0\},
  \\\nonumber
  &&
  v^{(i;kl)}=\delta_{k1} v^{(i;l)} +
  \delta_{k (l+n_1(i))} I_M ,\\\nonumber
  &&
  w^{(i;kl)}=\delta_{k 1} w^{(i;l)} +
  \delta_{k(l+n_2(i))} I_M  ,
  \end{eqnarray}
where  $n_0$ is a positive integer parameter, $n_1(i)$ and $n_2(i)$ are arbitrary positive  integer functions of positive integer argument,
$v^{(i;l)}$ and $w^{(i;l)}$ are $M\times M$ matrix fields.
In particular, if $n_1(i)=n_2(i)=1$, then this reduction becomes Frobenius one \cite{ZS_2}, which is shown in Fig.1. 
Eq.(\ref{G_F}) requires the following diagonal block-structures for  $ a$, $\nu$ and $\mu$:
\begin{eqnarray}\label{nu}
a={\mbox{diag}}(\underbrace{\tilde  a,\dots,\tilde  a}_{n_0}),\;\;\nu={\mbox{diag}}(\underbrace{\tilde \nu,\dots,\tilde \nu}_{n_0}), \;\;\mu={\mbox{diag}}(\underbrace{\tilde \mu,\dots,\tilde \mu}_{n_0}), \;\; 
\end{eqnarray}
($\tilde \nu$, $\tilde \mu$ and $\tilde  a$  are $M\times M$ diagonal constant matrices) and 
the appropriate  block-structures for the functions $\Psi^{(l)}$ and $\tilde \Lambda^{(l)}$:
\begin{eqnarray} &&
\Psi^{(l)}=\{\Psi^{(l;nm)},\;n,m=1,\dots,n_0\},\\\nonumber
&&
\tilde \Lambda^{(l)}={\mbox{diag}}(\tilde \Lambda^{(l;1)},\dots,\tilde\Lambda^{(l;n_0)}),\;\;l=1,\dots,2N_0.
\end{eqnarray}
Here $\Psi^{(l;nm)}$ are $M\times M$ matrix functions and 
$\tilde \Lambda^{(l;n)}$ are $M\times M$ diagonal matrix functions.
Eq.(\ref{b1}) reduces to the following one:
\begin{eqnarray}\label{Psi_ex}
\Psi^{(l;nm)}\tilde \Lambda^{(l;m)}&=&
\sum_{i=1}^{N_0}\sum_{j=1}^{n_0}\left[\Big(\delta_{n1}v^{(i;j)} +
\delta_{n(j+n_1(i))} \Big)
\Psi^{(l;jm)}_x +
\Big(\delta_{n1} w^{(i;j)} + \right.\\\nonumber
&&\left.
\delta_{n(j+n_2(i))}  \Big)
\Psi^{(l;jm)}\right](\tilde\Lambda^{(l;m)})^{-i+1},\\\nonumber
&&
l=1,\dots,2N_0\;\;n,m=1,\dots,n_0,
\end{eqnarray}
while eqs.(\ref{t},\ref{x}) and (\ref{nl_Lam_m},\ref{Lam_x}) yield respectively
\begin{eqnarray}\label{t_ex}
&&
\Psi^{(l;jm)}_{t_n} + \Psi^{(l;jm)}_{z_n} \tilde \Lambda^{(l;m)} =0,\\\label{x_ex}
&&
\Psi^{(l;jm)}_{xx}= \tilde a \Psi^{(l;jm)} \tilde \Lambda^{(m)}+\tilde \nu \Psi^{(l;jm)}_x + \tilde \mu\Psi^{(l;jm)},
\\\label{nl_Lam_m_ex}
&&
\tilde \Lambda^{(l;m)}_{t_n} + \tilde \Lambda^{(l;m)}_{z_n} \tilde \Lambda^{(l;m)} =0,\;\;\;\tilde \Lambda^{(l;m)}_x=0,\\\nonumber
&&
n=1,2,\dots,\;\;l=1,\dots,2N_0,\;\;j,m=1,\dots,n_0.
\end{eqnarray}
Then the chains of nonlinear PDEs (\ref{nl1},\ref{nl2},\ref{nl3},\ref{nl4}) get the following block structures:
\begin{eqnarray}
&&
E^{(mi)}_n=\{E^{(mi;l)}_n\delta_{k1},\;\;k,l,=1,\dots,n_0\}=0,
\\\nonumber
&&
\tilde E^{(mi)}=\{\tilde E^{(mi;l)}\delta_{k1},\;\;k,l,=1,\dots,n_0\}=0,\\\nonumber
&&m=0,1,\;\;\;i=1,\dots,N_0.
\end{eqnarray} 
where 
\begin{eqnarray}\label{nl1_22}
E^{(1i;l)}_n&:=&
v^{(i;l)}_{t_n} +v^{(1;1)}_{z_n} (w^{(i;l)} +v^{(i;l)}_x + v^{(i;l)}\tilde \nu )
+ 
(v^{(1;1)}_{z_n} v^{(1;1)}+
v^{(1;1+n_1(1))}_{z_n} )\tilde  a v^{(i;l)}
+\\\nonumber
&&v^{(1;1)}_{z_n} v^{(1;l+n_1(i))}\tilde a
+
w^{(1;1)}_{z_n} v^{(i;l)}+(Q_1^{(i;l)})_{z_n}=0,\\\label{nl2_22}
E^{(0i;l)}_n&:=&
w^{(i;l)}_{t_n}  + v^{(1;1)}_{z_n} (w^{(i;l)}_x+v^{(i;l)} \tilde \mu+v^{(i+1;l)}\tilde  a)
+
(v^{(1;1)}_{z_n}v^{(1;1)}+
v^{(1;1+n_1(1))}_{z_n})\tilde a
w^{(i;l)}+\\\nonumber
&&
v^{(1;1)}_{z_n}v^{(1;l+n_2(i))}\tilde a
+
w^{(1;1)}_{z_n} w^{(i;l)}+(Q_2^{(i;l)})_{z_n}=0,
\end{eqnarray}
\begin{eqnarray}\label{nl3_22}
\tilde E^{(1i;l)}&:=&
v^{(i;l)}_{xx} +[w^{(i;l)},\tilde \nu] + 2 w^{(i;l)}_x + 2 v^{(i;l)}_x \tilde \nu-\nu v^{(i;l)}_x +[v^{(i;l)},\tilde \mu] + [v^{(i;l)},\tilde \nu]\tilde \nu +\\\nonumber
&&
[v^{(1;1)},\tilde \nu]
\tilde  a v^{(i;l)}+[v^{(1;l+n_1(i))},\tilde \nu]
\tilde  a  +2 v^{(1;1)}_x \tilde  a v^{(i;l)}+2 v^{(1;l+n_1(i))}_x \tilde a+
[w^{(1;1)},\tilde a] v^{(i;l)} +\\\nonumber
&&
[v^{(1;1)},\tilde  a](
w^{(i;l)} +v^{(i;l)}_x + v^{(i;l)}\tilde \nu) 
+
([v^{(1;1)},\tilde  a] v^{(1;1)}+[v^{(1;1+n_1(1))},\tilde  a])
\tilde a v^{(i;l)}+\\\nonumber
&&
[v^{(1;1)},\tilde  a]v^{(1;l+n_1(i))}\tilde a  +
[Q_1^{(i;l)},\tilde a]=0,\\\label{nl4_22}
\tilde E^{(0i;l)}&:=&
w^{(i;l)}_{xx} + [w^{(i;l)},\mu] + [w^{(1;1)}, a] w^{(i;l)}+ 2 v^{(i;l)}_x\mu + [v^{(i;l)},\tilde\nu]\tilde\mu + 
[v^{(1;1)},\tilde\nu]\tilde  a w^{(i;l)}+\\\nonumber
&&
[v^{(1;l+n_2(i))},\tilde\nu]\tilde  a + 
2 v^{(1;1)}_x\tilde  a w^{(i;l)}+2 v^{(1;l+n_2(i))}_x\tilde  a -\tilde \nu w^{(i;l)}_x + \\\nonumber
&&[v^{(1;1)},\tilde  a](w^{(i;l)}_x + v^{(i;l)}\tilde \mu  +v^{(i+1;l)}\tilde  a) +
([v^{(1;1)},\tilde  a] v^{(1;1)}+[v^{(1;1+n_1(1))},\tilde  a])\tilde  a 
w^{(i;l)} +\\\nonumber
&&
[v^{(1;1)},\tilde  a] v^{(1;l+n_2(i))}\tilde  a 
 +
2 v^{(i+1;l)}_x\tilde  a +[v^{(i+1;l)},\tilde \nu]\tilde  a+
 [Q_2^{(i;l)},\tilde  a] =0,
\end{eqnarray}
\begin{eqnarray}
&&
Q_1^{(i;l)}=v^{(1;l+n_1(i)+n_1(1))} \tilde a + v^{(1;l+n_2(i))}+v^{(1;l+n_1(i))}\tilde\nu + w^{(1;l+n_1(i))}+v^{(i+1;l)},\\\nonumber
&&
Q_2^{(i;l)}=v^{(1;l+n_2(i)+n_1(1))} \tilde a+ v^{(1;l+n_1(i))}\tilde\mu+ v^{(1;l+n_1(i+1))}\tilde a+w^{(1;l+n_2(i))}+w^{(i+1;l)}. 
\end{eqnarray}
We see that after the reduction (\ref{G_F}) the chains of  nonlinear PDEs (\ref{nl1},\ref{nl2},\ref{nl3},\ref{nl4}) acquire one more discrete variable.

To write the complete system of nonlinear PDEs we, first of all,  put $i=l=1$ in the  chains of PDEs (\ref{nl1_22}-\ref{nl4_22}) and take two values of $n$ ($n=1,2$) in eqs.(\ref{nl1_22},\ref{nl2_22})
(remember, that deriving the complete system (\ref{nl1_1}-\ref{nl4_1}) in  Sec.\ref{Section:diff_reduction} we fixed  $n=1$ in eqs.(\ref{nl1},\ref{nl2})):
\begin{eqnarray}\label{nl1_2}
E^{(11;1)}_n&:=&
v^{(1;1)}_{t_n} +v^{(1;1)}_{z_n} (w^{(1;1)} +v^{(1;1)}_x + v^{(1;1)}\tilde \nu )
+ (v^{(1;1)}_{z_n} v^{(1;1)}+\\\nonumber
&&
v^{(1;1+n_1(1))}_{z_n} )\tilde  a v^{(1;1)}
+v^{(1;1)}_{z_n} v^{(1;1+n_1(1))}\tilde  a
+
w^{(1;1)}_{z_n} v^{(1;1)}+
(Q_1^{(1;1)})_{z_n}=0,\\\label{nl2_2}
E^{(01;1)}_n&:=&
w^{(1;1)}_{t_n}  + v^{(1;1)}_{z_n} (w^{(1;1)}_x+v^{(1;1)} \tilde \mu+v^{(2;1)}\tilde a)
+
(v^{(1;1)}_{z_n}v^{(1;1)}+\\\nonumber
&&
v^{(1;1+n_1(1))}_{z_n})\tilde a
w^{(1;1)}+
v^{(1;1)}_{z_n}v^{(1;1+n_2(1))}\tilde a
+
w^{(1;1)}_{z_n} w^{(1;1)}+(Q_2^{(1;1)})_{z_n}=0,
\end{eqnarray}
\begin{eqnarray}\label{nl3_2}
\tilde E^{(11;1)}&:=&
v^{(1;1)}_{xx} +[w^{(1;1)},\tilde \nu] + 2 w^{(1;1)}_x + 2 v^{(1;1)}_x \tilde \nu-\tilde \nu v^{(1;1)}_x +[v^{(1;1)},\tilde \mu] + [v^{(1;1)},\tilde \nu]\tilde \nu +\\\nonumber
&&
[v^{(1;1)},\tilde \nu]
\tilde  a v^{(1;1)}+[v^{(1;1+n_1(1))},\tilde \nu]
\tilde  a  +2 v^{(1;1)}_x \tilde  a v^{(1;1)}+2 v^{(1;1+n_1(1))}_x\tilde  a +\\\nonumber
&&
[w^{(1;1)},\tilde  a] v^{(1;1)}+
[v^{(1;1)},\tilde  a](
w^{(1;1)} +v^{(1;1)}_x + v^{(1;1)}\tilde \nu) 
+\\\nonumber
&&
([v^{(1;1)},\tilde  a]v^{(1;1)}+[v^{(1;1+n_1(1))},\tilde  a])
\tilde a v^{(1;1)}+
[v^{(1;1)},\tilde  a]v^{(1;1+n_1(1))}\tilde a +[Q_1^{(1;1)},\tilde a]=0,\\\label{nl4_2}
\tilde E^{(01;1)}&:=&
w^{(1;1)}_{xx} + [w^{(1;1)},\tilde \mu] + [w^{(1;1)}, \tilde a] w^{(1;1)} + 2 v^{(1;1)}_x\tilde \mu +\\\nonumber
&& [v^{(1;1)},\tilde\nu]\tilde\mu + 
[v^{(1;1)},\tilde\nu]\tilde  a w^{(1;1)}+ 
[v^{(1;1+n_2(1))},\tilde\nu]\tilde  a + \\\nonumber
&&
2 v^{(1;1)}_x\tilde  a w^{(1;1)}+2 v^{(1;1+n_2(1))}_x\tilde  a -\tilde \nu w^{(1;1)}_x +[v^{(1;1)},\tilde  a](w^{(1;1)}_x + v^{(1;1)}\tilde \mu  +v^{(2;1)}\tilde  a) +\\\nonumber
&&
([v^{(1;1)},\tilde  a] v^{(1;1)}+
[v^{(1;1+n_1(1))},\tilde  a])\tilde  a 
w^{(1;1)} +[v^{(1;1)},\tilde  a] v^{(1;1+n_2(1))}\tilde  a 
 +
2 v^{(2;1)}_x\tilde  a +\\\nonumber
&& [v^{(2;1)},\tilde \nu]\tilde  a+[Q_2^{(1;1)},\tilde a] =0,
\end{eqnarray}
where
\begin{eqnarray}
 &&
Q_1^{(1;1)}=v^{(1;1+2n_1(1))} \tilde a + v^{(1;1+n_2(1))}+v^{(1;1+n_1(1))}\tilde\nu + w^{(1;1+n_1(1))}+v^{(2;1)},\\\nonumber
&&
Q_2^{(1;1)}=v^{(1;1+n_1(1)+n_2(1))} \tilde a+ v^{(1;1+n_1(1))}\tilde\mu+ v^{(1;1+n_1(2))}\tilde a+w^{(1;1+n_2(1))}+w^{(2;1)}.
\end{eqnarray}
One can eliminate $Q_1^{(1;1)}$ from the system (\ref{nl1_2},\ref{nl3_2}) resulting in the following complete system of two PDEs for the matrix fields $u=v^{(1;1)}$ and  $q= w^{(1;1)}+v^{(1;1+n_1(1))}$:
\begin{eqnarray}\label{NL1}
&&
(E^{(11;1)}_1)_{z_2} -(E^{(11;1)}_2)_{z_1}=0,
\\\label{NL3}
&&    
[E^{(11;1)}_1,\tilde a] -(\tilde E^{(11;1)})_{z_1}=0.
\end{eqnarray}
Three equations 
(\ref{nl2_2},\ref{nl4_2}) are not important because they introduce three  more fields ($w^{(1;1)}$, $Q^{(1;1)}_2$ and $p=v^{(2;1)}+v^{(1;1+n_2(1))}$) which do not appear in the system (\ref{NL1},\ref{NL3}).
In particular, if $\tilde  a$ is a scalar, eq.(\ref{NL3}
) must be replaced by the following one:
\begin{eqnarray}
\tilde E^{(11;1)}
=0.
\end{eqnarray}

Consider the scalar case ($M=1$). Then  only field  $u=v^{(1;1)}$ remains in the eq. (\ref{NL1}) which becomes eq.(\ref{intr_v}). Eq. (\ref{NL3}) is not important in this case.
Thus the system (\ref{NL1},\ref{NL3}) may be considered as a matrix generalization of eq.(\ref{intr_v}). Its integrability must be studied more carefully.

All in all, we have the following chain of transformations relating 
four systems of nonlinear  PDEs having different integrability properties (compare with Fig.1):
\begin{eqnarray}
\text{eq.}(\ref{nl_W})\stackrel{\text{eq.}(\ref{ww})}\longrightarrow 
\text{eq.}(\ref{SDYM})\stackrel{\text{eq.}(\ref{ww_i})}\longrightarrow 
\text{eqs.}(\ref{nl1_1}-\ref{nl4_1})\stackrel{\text{eqs.}(\ref{G_F})}\longrightarrow 
\text{eqs.}(\ref{NL1},\ref{NL3})\stackrel{\text{scalar case }}\longrightarrow \text {eq.}(\ref{intr_v}).
\end{eqnarray}

\section{Solutions to the nonlinear PDEs
 }
\label{Section:Solutions}

\subsection{Solutions to the system of nonlinear PDEs (\ref{nl1_1}-\ref{nl4_1})}
Solutions to the system of matrix  nonlinear PDEs (\ref{nl1_1}-\ref{nl4_1})  
may be written in terms of the functions $\Psi^{(m)}$ and $\tilde \Lambda^{(m)}$, $m=1,\dots,2N_0$, taken as  solutions to the linear systems (\ref{t},\ref{x}) and (\ref{nl_Lam_m},\ref{Lam_x}) respectively with $n=1$. 
They read:
\begin{eqnarray} \label{Psi_sol0}
\Psi^{(l)}_{\alpha\beta}(\vec{x}) &=&  \sum_{i=1}^2 \psi^{(l;i)}_{\alpha\beta}(z_1 -\tilde \Lambda^{(l)}_\beta t_1) e^{k^{(l;i)}_{\alpha\beta} x },\\\label{Lam_sol}
\tilde \Lambda^{(l)}_{\beta} &=& E^{(l)}_{\beta}(z_1 -\tilde \Lambda^{(l)}_\beta t_1) 
,\;\;\;
\alpha,\beta=1,\dots,n_0M,
\;\;l=1,\dots,2 N_0.
\end{eqnarray}
where $\psi^{(l)}_{\alpha\beta}(y)$ and  
$E^{(l)}_\beta (y)$ are arbitrary scalar functions of single scalar variable,  $\vec x=(x,z_1,t_1)$ is the list of all independent variables of the nonlinear PDEs,  $k^{(l;i)}_{\alpha\beta}$ are the roots of the  characteristics equation assotiated with eq.(\ref{g_r2}):
\begin{eqnarray}\label{char}
&&
k^2  - a_\alpha\Lambda^{(l)}_{\beta}-\nu_\alpha k -\mu_\alpha=0\;\;\Rightarrow\\\nonumber
&&
k^{(l;1)}_{\alpha\beta}=\frac{1}{2}\left(\nu_\alpha +\sqrt{(\nu_\alpha)^2+4  a \tilde\Lambda^{(l)}_\beta + 4\mu_\alpha}\right),\;\;
k^{(l;2)}_{\alpha\beta}=\frac{1}{2}\left(\nu_\alpha -\sqrt{(\nu_\alpha)^2+4  a \tilde\Lambda^{(l)}_\beta + 4\mu_\alpha}\right)
.
\end{eqnarray}

Now, we can use eq.(\ref{b1}) in order to find $v^{(i)}$ and $w^{(i)}$. The simplest nontrivial case corresponds to 
$N_0=2$. Then  eq.(\ref{b1}) reduces to the following four  matrix equations:
\begin{eqnarray}\label{Psi_Sato}
\Psi^{(m)}\tilde \Lambda^{(m)} =
\sum_{i=1}^4 \Big(v^{(i)} \Psi^{(m)}_x +w^{(i)} \Psi^{(m)}\Big) \Big(\tilde\Lambda^{(m)}\Big)^{-i+1},\;\;m=1,2,3,4.
\end{eqnarray}
These equations, in general, are uniquely solvable  for the matrix fields $v^{(i)}$ and $w^{(i)}$, $i=1,2$.  We do not represent the explicite expressions for them. 

\subsubsection{Dimensionality of the solution space}
\label{Section:dim1}
To define restrictions, generated by our algorithm, on the
 solution space to eqs.(\ref{nl1_1}-\ref{nl4_1})  we, first of all, 
write equation (\ref{b1}) in the  following compact form:
\begin{eqnarray}\label{comp}
\Psi^{(m)}\tilde \Lambda^{(m)} = \sum_{n=1}^{N_0}\vec V^{(n)} \hat \Pi^{(nm)},\;\;m=1,\dots,2N_0,
\end{eqnarray}
where $\vec V^{(n)}= (v^{(n)}\;\;w^{(n)})$ and $\hat \Pi^{(nm)}$ is the following  $2N_0 n_0 M \times 2N_0 n_0 M$ invertible operator:
\begin{eqnarray}\label{Pi}
\hat \Pi^{(nm)}=\left(
\begin{array}{c}
\Pi^{(nm)}_x\cr
\Pi^{(nm)}
\end{array}\right),\;\;\Pi^{(nm)}=\Psi^{(m)}\Lambda^{(-n+1)},\;\;n=1,\dots,N_0,\;\;m=1,\dots,2N_0.
\end{eqnarray}
Introduce operator $\tilde \Pi^{(mj)}$ by the formula
\begin{eqnarray}
\sum_{m=1}^{2 N_0}\hat \Pi^{(nm)}\tilde \Pi^{(mj)}=\delta_{nj}I_{n_0M},\;\;
\tilde \Pi^{(mn)}=(\tilde \Pi^{(mn)}_1\;\;\tilde \Pi^{(mn)}_2),\;\;n,j=1,\dots,N_0.
\end{eqnarray}
Then eq.(\ref{comp}) yields
\begin{eqnarray}\label{vw}
v^{(n)}=\sum_{m=1}^{2N_0}\Psi^{(m)}\tilde \Lambda^{(m)} \tilde \Pi^{(mn)}_1,\;\;
w^{(n)}=\sum_{m=1}^{2N_0}\Psi^{(m)}\tilde \Lambda^{(m)} \tilde \Pi^{(mn)}_2,\;\;n=1,\dots,N_0.
\end{eqnarray}

Now let us turn to the eq.(\ref{Psi_sol0}). We represent arbitrary functions 
$\psi^{(l;i)}_{\alpha\beta}(y)$ in the following integral form:
\begin{eqnarray}
\psi^{(l;i)}_{\alpha\beta}(y)=\int d q \hat \psi^{(l;i)}_{\alpha\beta}(q) e^{q y},\;\;i=1,2,\;\;l=1,\dots,2N_0,\;\;\alpha,\beta=1,\dots,n_0 M
\end{eqnarray}
where $\hat \psi^{(l;i)}_{\alpha\beta}(q)$ are arbitrary functions of argument and one integrates  over the whole space of the parameter $q$ which  is complex in general.
Then eq.(\ref{Psi_sol0}) may be written in the following form:
\begin{eqnarray}\label{Psi_l_F}
\Psi^{(l)}_{\alpha\beta}(\vec{x}) &=&  \sum_{i=1}^2 \int dq
\hat \psi^{(l;i)}_{\alpha\beta}(q) e^{q z_1-\tilde \Lambda^{(l)}_\beta q t_1+k^{(l;i)}_{\alpha\beta} x},\;\;l=1,\dots,2N_0,\;\;\alpha\beta=1,\dots,n_0 M.
\end{eqnarray}
Substituting expression (\ref{Psi_l_F}) for $\Psi^{(m)}$ into 
eqs.(\ref{vw}) we write them as follows:
\begin{eqnarray}\label{v_j}
&&
v^{(j)}_{\alpha\beta}(\vec x)= \sum_{\gamma=1}^{n_0M} \sum_{m=1}^{2N_0}\sum_{i=1}^2 \int dq
\hat \psi^{(m;i)}_{\alpha\gamma}(q)\tilde \Lambda^{(m)} e^{q z_1-\tilde \Lambda^{(m)}_\gamma q t_1+k^{(m;i)}_{\alpha\gamma} x}(\tilde \Pi^{(mj)}_1(\vec x))_{\gamma\beta},\\\nonumber
&&
w^{(j)}_{\alpha\beta}(\vec x)= \sum_{\gamma=1}^{n_0M} \sum_{m=1}^{2N_0}\sum_{i=1}^2 \int dq
\hat \psi^{(m;i)}_{\alpha\gamma}(q)\tilde \Lambda^{(m)} e^{q z_1-\tilde \Lambda^{(m)}_\gamma q t_1+k^{(m;i)}_{\alpha\gamma} x}(\tilde \Pi^{(mj)}_2(\vec x))_{\gamma\beta},\\\nonumber
&&j=1,\dots,N_0,\;\;\alpha,\beta=1,\dots,n_0M.
\end{eqnarray}
These formulae might be considered as integral representations of $v^{(j)}$ and $w^{(j)}$  if $\tilde \Pi^{(mj)}_n$ ($n=1,2$) would not depend on $\vec x$. However, $\tilde \Pi^{(mj)}$ do depend on $\vec x$, so that eq.(\ref{v_j}) has more complicated sense. Nevertheless, in order to estimate the  dimensionality  of the solution space we consider eqs.(\ref{v_j}) as integral representations of $v^{(j)}$ and $w^{(j)}$ with the kernel
\begin{eqnarray}\label{kernel1}
R^{(i)}_{\alpha\gamma}(x,z_1,t_1; q,\Lambda^{(m)}_\gamma)=e^{q z_1-\tilde \Lambda^{(m)}_\gamma q t_1+k^{(m;i)}_{\alpha\gamma} x},\;\;i=1,2.
\end{eqnarray} 
Formulae (\ref{v_j})   transform functions  $v^{(j)}_{\alpha\beta}$ and $w^{(j)}_{\alpha\beta}$ of three continues variables $x$, $z_1$ and $t_1$ into  functions  $\hat \psi^{(m;i)}_{\alpha\beta}(q)$  ($i=1,2$)  depending on continues variable $q$ and discrete variable $m$. Discrete variable $m$ is assotiated with $\tilde \Lambda^{(m)}$ in the kernel of the integral transformation, where $m=1,\dots,2N_0$ and $N_0$ is an arbitrary positive integer. 
Thus, we may state that the solution space to the eqs.(\ref{nl1_1}-\ref{nl4_1}) has freedom of two arbitrary  functions, $\psi^{(m;1)}_{\alpha\beta}(q)$ and $\psi^{(m;2)}_{\alpha\beta}(q)$,  of one continues and one discrete variable, i.e. one has two-dimensional solution space. Using these  functions we may approximate (at least formally) two initial conditions, for instance, 
\begin{eqnarray}
(v^{(1)}_{\alpha\beta}|_{t_1=0},w^{(1)}_{\alpha\beta}|_{t_1=0}) \; \to \;(\psi^{(m;1)}_{\alpha\beta}(q), \psi^{(m;2)}_{\alpha\beta}(q)),
\end{eqnarray}
and, as a consequence, we are able to solve the initial value problem (IVP) for the system (\ref{nl1_1}-\ref{nl4_1}) (which describes evolution of $v^{(1)}$ and $w^{(1)}$) with "approximate" initial conditions. We say "approximate initial conditionÙ" since functions $\psi^{(m;i)}_{\alpha\beta}(q)$ ($i=1,2$) have one continues and one discrete variable, while both variables must be continues in order to represent arbitrary initial conditions precisely.
 We conclude that  one  provides the same variety of  solutions   to eqs.(\ref{nl1_1}-\ref{nl4_1}) as Sato approach does to the classical $S$-integrable PDEs \cite{OSTT}.

 \subsection{Solutions to the system (\ref{NL1},\ref{NL3}) and to its scalar version (\ref{intr_v})}  

Eqs.(\ref{Psi_ex}) have been derived as an algebraic system for the fields $v^{(i;j)}$ and $w^{(i;j)}$ in this case. We split the system (\ref{Psi_ex}) into two subsystems.
First one corresponds to $n=1$:
\begin{eqnarray}\label{Psi1}
\Psi^{(l;1m)}\tilde \Lambda^{(l;m)}&=&
\sum_{i=1}^{N_0}\sum_{j=1}^{n_0}\left[v^{(i;j)} 
\Psi^{(l;jm)}_x +
\ w^{(i;j)} 
\Psi^{(l;jm)}\right](\tilde \Lambda^{(l;m)})^{-i+1},\\\nonumber
&&
l=1,\dots,2 N_0\;\;,\;\;m=1,\dots,n_0.
\end{eqnarray}
This is the system of $2 N_0 n_0$  linear algebraic $M\times M$ matrix equations for the same number of matrix  fields $v^{(i;l)}$ and $w^{(i;l)}$, $i=1,\dots,N_0$, $j=1,\dots,n_0$. Namely eqs.(\ref{Psi1})  yield functions $u=v^{(1;1)}$ and  $q=w^{(1;1)}+v^{(1;1+n_1(1))}$
as solution to the system (\ref{NL1},\ref{NL3}).
Second subsystem corresponds to $n>1$ in (\ref{Psi_ex}):
\begin{eqnarray}\label{Psi_nm}
\Psi^{(l;nm)}\tilde \Lambda^{(l;m)}&=&
\sum_{i=0}^{N_0}\left[
\Psi^{(l;(n-n_1(i))m)}_x +
\Psi^{(l;(n-n_2(i))m)}\right](\tilde \Lambda^{(l;m)})^{-i+1},\\\nonumber
\Psi^{(l;ij)}&=&0,\; {\mbox{if}},\;i\le 0,\\\nonumber
&&
l=1,\dots,2 N_0\;\;,\;\;n,m=1,\dots,n_0,
\end{eqnarray}
This equation expresses recursively the functions
 $\Psi^{(l;nm)}$, $n>1$, in terms of the functions 
 $\Psi^{(l;1m)}$ and their $x$-derivatives. The simplest case corresponds to $n_j(i)=1$, $\forall i,j$ (Frobenius reduction).
 
 Functions $\Psi^{(l;nm)}$ and $\tilde\Lambda^{(l;m)}$ are solutions to the system (\ref{t_ex}-\ref{nl_Lam_m_ex})  with $n=1,2$ (compare with  eqs.(\ref{Psi_sol0},\ref{Lam_sol})):  
\begin{eqnarray} \label{Psi_sol}
\Psi^{(l;1m)}_{\alpha\beta}(\vec{x}) &=&\sum_{i=1}^2 \psi^{(lm;i)}_{\alpha\beta}(z_1 -\tilde\Lambda^{(l;m)}_\beta t_1,z_2 -\tilde\Lambda^{(l;m)}_\beta t_2) 
e^{k^{(lm;i)}_{\alpha\beta} x }
,\\\label{Lam_sol_ex}
\tilde \Lambda^{(l;m)}_{\beta} &=& E^{(lm)}_{\beta}(z_1 -\tilde \Lambda^{(l;m)}_\beta t_1,z_2 -\tilde\Lambda^{(l;m)}_\beta t_2), 
\\\nonumber
&&
k^{(lm;1)}_{\alpha\beta}=\frac{1}{2}\left(\tilde \nu_\alpha +\sqrt{(\tilde \nu_\alpha)^2+4 \tilde  a \tilde \Lambda^{(l;m)}_\beta + 4\tilde \mu_\alpha}\right),\\\nonumber
&&
k^{(lm;2)}_{\alpha\beta}=\frac{1}{2}\left(\tilde \nu_\alpha -\sqrt{(\tilde \nu_\alpha)^2+4 \tilde  a \tilde \Lambda^{(l;m)}_\beta + 4\tilde \mu_\alpha}\right),
\\\nonumber
&&m=1,\dots,n_0,\;\;l=1,\dots,2N_0,\;\;\alpha,\beta=1,\dots,M.
\end{eqnarray}
Here $\psi^{(lm;i)}_{\alpha\beta}(y_1,y_2)$ and  $E^{(lm)}_\beta(y_1,y_2)$ are arbitrary scalar functions of two scalar  arguments. 

Since the functions $\Psi^{(l;1m)}$ depend explicitely  on the functions $\tilde \Lambda^{(l;m)}$, the functions $v^{(i;j)}$ and $w^{(i;j)}$  depend explicitely on $\tilde \Lambda^{(l;m)}$ as well. Functions $\tilde \Lambda^{(l;m)}$ describe the break of the wave profiles because they are solutions of eqs.(\ref{Lam_sol_ex}). 
Consequently,  the functions $v^{(i;j)}$ and $w^{(i;j)}$  (in
  particular, solutions to the nonlinear PDEs 
  (\ref{NL1},\ref{NL3})) exhibit the break of the wave profiles 
  as well, unless $\tilde\Lambda^{(l;m)}=const$ $\forall l,m$. One gets explicite solutions in
  the later case.

Consider the scalar case corresponding to eq.(\ref{intr_v}), i.e. $M=1$
 and diagonal  matrices  $\tilde a$, $\tilde \nu$, 
$\tilde \mu$ and $\tilde \Lambda^{(l;m)}$ become scalars. 
Eqs.(\ref{Psi_sol},\ref{Lam_sol_ex}) read 
\begin{eqnarray} \label{Psi_sol_sc}
&&
\Psi^{(l;1m)}(\vec{x}) = \sum_{i=1}^2\psi^{(lm;i)}(z_1 -\tilde\Lambda^{(l;m)} t_1,z_2 -\tilde\Lambda^{(l;m)} t_2) 
e^{k^{(lm;i)} x }
\\\label{Lam_sol_ex_sc}
&&
\tilde \Lambda^{(l;m)} = E^{(lm)}(z_1 -\tilde \Lambda^{(l;m)} t_1,z_2 -\tilde\Lambda^{(l;m)} t_2)
,\\\nonumber
&&k^{(lm;1)}=\frac{1}{2}\left(\tilde \nu 
+\sqrt{\tilde \nu^2+4  \tilde a \tilde \Lambda^{(l;m)} + 4\tilde \mu}\right),\;\;k^{(lm;1)}=\frac{1}{2}\left(\tilde \nu 
-\sqrt{\tilde \nu^2+4  \tilde a \tilde \Lambda^{(l;m)} + 4\tilde \mu}\right),
\\\nonumber
&&
m=1,\dots,n_0,\;\;l=1,\dots,2N_0.
\end{eqnarray}
Reduction (\ref{red_I}) corresponding to the eq.(\ref{h5_red_I})
reduces the eq.(\ref{Psi_sol_sc}) into the following one:
 \begin{eqnarray} \label{Psi_sol_sc_red}
\Psi^{(l;1m)}(\vec{x}) &=& \sum_{i=1}^2\psi^{(lm;i)}
e^{k^{(lm;i)} (x  -\tilde\Lambda^{(l;m)} y-(\tilde\Lambda^{(l;m)})^2 t)}
,\\\nonumber
&&
m=1,\dots,n_0, \;\;l=1,\dots,2N_0,
\end{eqnarray}
where $\psi^{(l;m)}$ and $\tilde \Lambda^{(l;m)}$ are arbitrary scalar constant parameters. Emphasise, that $\tilde \Lambda^{(l;m)}$ does not depend on variables $\vec x$ and  there is no arbitrary functions in the available solution manifold. 

\subsubsection{Dimensionality of the solution space}
Let us define restrictions, generated by our algorithm, on the   solution space to the nonlinear PDEs (\ref{NL1},\ref{NL3}). 
First of all we represent 
the formal solution to the system (\ref{Psi_nm})  as follows:
\begin{eqnarray}\label{first}
\Psi^{(l;nm)} = \Psi^{(l;1m)} P^{(lmn)},\;\;\;P^{(lm1)}\equiv I_M,\;\;l=1,\dots,2 N_0,\;\;n,m=1,\dots,n_0,
\end{eqnarray}
reflecting the fact that eqs.(\ref{Psi_nm}) must express $\Psi^{(l;nm)}$ ($n>1$) in terms of $\Psi^{(l;1m)}$, $\forall\;l,m$.  
Now we may represent eq.(\ref{Psi1}) in the following compact form:
\begin{eqnarray}\label{vvww}
\Psi^{(l;1m)}\Lambda^{(l;m)} &=& \sum_{i=1}^{N_0}\sum_{j=1}^{n_0}
v^{(i;j)} \Pi^{(ij;lm)}_x +w^{(i;j)} \Pi^{(ij;lm)}=\sum_{i=1}^{N_0}\sum_{j=1}^{n_0}\vec V^{(i;j)} \hat \Pi^{(ij;lm)},\\\nonumber
&&
l=1,\dots,2 N_0,\;\;m=1,\dots,n_0,
\end{eqnarray}
where 
\begin{eqnarray}\label{Psi_lm}
&&
\vec V^{(i;j)}= (v^{(i;j)}\;\;w^{(i;j)}),\;\;\hat \Pi^{(ij;lm)}=\left(
\begin{array}{c}
\Pi^{(ij;lm)}_x \cr
\Pi^{(ij;lm)}
\end{array}
\right),\;\;
\Pi^{(ij;lm)}=\Psi^{(l;1m)} P^{lmj}(\Lambda^{(l;m)})^{-i+1},\\\nonumber
&&
i=1,\dots,N_0, \;\;j,m=1,\dots,n_0,\;\;l=1,\dots, 2N_0
.
\end{eqnarray}
 Introduce operator $\tilde \Pi^{(ml;ji)}$ by the following formula:
 \begin{eqnarray}
&&
\sum_{l=1}^{2N_0}\sum_{m=1}^{n_0} \hat \Pi^{(ij;lm)} \tilde \Pi^{(ml;kn)}=\delta_{in}\delta_{jk}I_M,\;\;
 \tilde \Pi^{(ml;kn)}=\left(
\begin{array}{cc}
\tilde \Pi^{(ml;kn)}_1 &
\tilde \Pi^{(ml;kn)}_2
\end{array}
\right),\\\nonumber
&&
i,n=1,\dots,N_0, \;\;j,k=1,\dots,n_0.
 \end{eqnarray}
Then eq.(\ref{vvww}) yields:
\begin{eqnarray}\label{vw_lm}
&&
v^{(n;j)}=\sum_{m=1}^{2N_0}\sum_{l=1}^{n_0}\Psi^{(l;1m)}\tilde \Lambda^{(l;m)} \tilde \Pi^{(ml;jn)}_1,\;\;
w^{(n;j)}=\sum_{m=1}^{2N_0}\sum_{l=1}^{n_0}\Psi^{(l;1m)}\tilde \Lambda^{(l;m)} \tilde \Pi^{(ml;jn)}_2,\\\nonumber
&&
n=1,\dots,N_0,\;\;j=1,\dots,n_0,
\end{eqnarray}

Now let us turn to eq. (\ref{Psi_sol}). We  represent arbitrary functions 
$\psi^{(lm;i)}_{\alpha\beta}(y_1,y_2)$ in the following integral  form 
 \begin{eqnarray}
 &&
 \psi^{(lm;i)}_{\alpha\beta}(y_1,y_2)=\int d q_1 dq_2 \hat\psi^{(lm;i)}_{\alpha\beta}(q_1,q_2) e^{ q_1 y_1+ q_2 y_2},\\\nonumber
 &&
 l=1,\dots,2N_0,\;\;m=1,\dots,n_0,\;\;\alpha,\beta=1,\dots,M,
 \end{eqnarray}
 where $\hat\psi^{(lm;i)}_{\alpha\beta}(q_1,q_2)$ are arbitrary functions of two arguments and one integrates over the whole two dimensional space of variables $q_1$ and $q_2$ which are complex in general.  
 Then
  eq.(\ref{Psi_sol}) may be written as follows:
 \begin{eqnarray}\label{Psi_lm_F}
 \Psi^{(l;1m)}_{\alpha\beta}(\vec{x}) &=&\sum_{i=1}^2 \int dq_1dq_2\hat\psi^{(lm;i)}_{\alpha\beta}(q_1,q_2) e^{ q_1 z_1+ q_2 z_2 -\tilde\Lambda^{(l;m)}_\beta(q_1 t_1 + q_2 t_2)+k^{(lm;i)}_{\alpha\beta} x },\\\nonumber
 &&
 l=1,\dots,2N_0,\;\;m=1,\dots,n_0,\;\;\alpha,\beta=1,\dots,M.
 \end{eqnarray} 
Substituting expression (\ref{Psi_lm_F}) for $\Psi^{(l;1m)}_{\alpha\gamma}$ into 
eqs.(\ref{vw_lm}) we write them in the following form:
\begin{eqnarray}\label{v_ij}
&&
v^{(n;j)}_{\alpha\beta}(\vec x)= \\\nonumber
&&
\sum_{\gamma=1}^M \sum_{m=1}^{2N_0}\sum_{l=1}^{n_0}\sum_{i=1}^2 \int dq_1dq_2
\hat \psi^{(lm;i)}_{\alpha\gamma}(q_1,q_2) e^{ q_1 z_1+ q_2 z_2 -\tilde\Lambda^{(l;m)}_\gamma(q_1 t_1 + q_2 t_2)+k^{(lm;i)}_{\alpha\gamma} x }(\tilde \Pi^{(ml;jn)}_1(\vec x))_{\gamma\beta},\\\nonumber
&&
w^{(n;j)}_{\alpha\beta}(\vec x)= \\\nonumber
&&
\sum_{\gamma=1}^M \sum_{m=1}^{2N_0}\sum_{l=1}^{n_0}\sum_{i=1}^2 \int dq_1dq_2
\hat \psi^{(lm;i)}_{\alpha\gamma}(q_1,q_2) e^{ q_1 z_1+ q_2 z_2 -\tilde\Lambda^{(l;m)}_\gamma(q_1 t_1 + q_2 t_2)+k^{(lm;i)}_{\alpha\gamma} x }(\tilde \Pi^{(ml;jn)}_2(\vec x))_{\gamma\beta},\\\nonumber
&&
n=1,\dots,N_0,\;\;j=1,\dots,n_0.
\end{eqnarray}

Now we show that the solution space to the five-dimensional matrix eqs.(\ref{NL1},\ref{NL3}) has two arbitrary $M\times M$ matrix functions of two continues and one discrete variable
(compare with Sec.\ref{Section:dim1}).
For this purpose we consider
formulae (\ref{v_ij}) as integral representations of $v^{(n;j)}$ and $w^{(n;j)}$     with the kernel
\begin{eqnarray}\label{kernel2}
R^{(i)}_{\alpha\gamma}(x,z_1,z_2,t_1,t_2; q_1,q_2,\Lambda^{(l;m)}_\gamma)=e^{q_1 z_1+q_2 z_2 -\tilde\Lambda^{(l;m)}_\gamma(q_1 t_1 + q_2 t_2)+k^{(lm;i)}_{\alpha\gamma} x },\;\;i=1,2.
\end{eqnarray} 
Formulae (\ref{v_ij})   transform functions  $v^{(n;j)}_{\alpha\beta}$ and $w^{(n;j)}_{\alpha\beta}$ of five continues variables $x$, $z_i$ and $t_i$ ($i=1,2$) into  functions  $\hat \psi^{(lm;i)}_{\alpha\beta}(q_1,q_2)$  ($i=1,2$)  depending on two continues variables $q_1$, $q_2$ and two discrete variables $l,m$. However, two discrete variables $l$ and $m$ are assotiated with  single discrete  variable $\tilde \Lambda^{(l;m)}$ in the kernel of the integral transformation, where $l=1,\dots,2N_0$, $m=1,\dots,n_0$  and $N_0$, $n_0$ are  arbitrary positive integers. 
Thus, we may state that the solution space to the eqs.(\ref{NL1},\ref{NL3}) (describing the evolution of $v^{(1;1)}$) has freedom of two arbitrary functions, $\hat \psi^{(lm;1)}_{\alpha\beta}(q_1,q_2)$ and $\hat \psi^{(lm;2)}_{\alpha\beta}(q_1,q_2)$,  of two continues  and one discrete variable, i.e. one has three-dimensional solution space with two arbitrary functions of three variable. Note that this freedom is not enough in order to solve, for instance, the IVP  for the matrix system (\ref{NL1},\ref{NL3}) even with "approximate initial conditions" like it was done in Sec.\ref{Section:dim1}. In fact, in order to approximate arbitrary initial condition one needs an (infinitely) big number  of such arbitrary functions of three variables. Thus, further modifications of our algorithm with the purpose to increase the dimensionality of the solution space to the derived nonlinear PDEs  becomes an important problem to be resolved.

Solution space to the scalar equation (\ref{intr_v})  has the same dimensionality so that formulae (\ref{first}-\ref{kernel2})  remains valid with $M=1$. 

Consider reduction (\ref{red_I}) and estimate the dimensionality of the available solution space to the (2+1)-dimensional PDE (\ref{h5_red_I}). As we have seen in both this subsection and  Sec.\ref{Section:dim1} the dimensionality of the solution space is completely defined by the kernel of the integral transformation, see eqs.(\ref{kernel1}) and (\ref{kernel2}). Reduction (\ref{red_I}) reduces eq.(\ref{kernel2}) with $M=1$  to the following one
\begin{eqnarray}\label{kernel3}
R^{(i)}(x,y,t; \Lambda^{(l;m)})=e^{
 k^{(lm;i)} (x  -\tilde\Lambda^{(l;m)} y-(\tilde\Lambda^{(l;m)})^2 t)}, \;\;i=1,2.
\end{eqnarray} 
This kernel has one discrete parameter $\Lambda^{(l;m)}$ and, as a consequence, the available solution space is one-dimensional with two arbitrary functions of single discrete variable. Number of arbitrary functions is predicted by two different values of the superscript $i$ in eq.(\ref{kernel3}).

\subsubsection{Simplest example of the solution to eq.(\ref{intr_v})}
The simplest nontrivial example of the solution to  eq.(\ref{intr_v}) is assotiated with  $N_0=n_0=1$. Then eq.(\ref{Psi_sol_sc}) reads
\begin{eqnarray} \label{Psi_sol_sc_ex}
\Psi^{(l;11)}(\vec{x}) &=&\sum_{i=1}^2 \psi^{(l1;i)}(z_1 -\tilde\Lambda^{(l;1)} t_1,z_2 -\tilde\Lambda^{(l;1)} t_2) 
e^{k^{(l1;i)} x }
,\;\;l=1,2,
\end{eqnarray}
  where $\tilde \Lambda^{(l;1)}$ ($l=1,2$) are solutions to the nonlinear  algebraic equations (\ref{Lam_sol_ex_sc}) which reduce to the following ones:
\begin{eqnarray}
\label{nl_Lam_ex}
\tilde \Lambda^{(l;1)} &=& E^{(l1)}(z_1 -\tilde \Lambda^{(l;1)} t_1,z_2 -\tilde\Lambda^{(l;1)} t_2),\;\;l=1,2.
\\\nonumber
\end{eqnarray} 
Then eqs.(\ref{Psi1}) become the system of  two following equations:
\begin{eqnarray}\label{Psi1_ex}
\Psi^{(l;11)}\tilde \Lambda^{(l;1)}&=&
v^{(1;1)} 
\Psi^{(l;11)}_x +
\ w^{(1;1)} 
\Psi^{(l;11)},\;\;\;
l=1,2 .
\end{eqnarray}
It has the following solution:
\begin{eqnarray}\label{solution}
u&\equiv& v^{(1;1)}=\frac{\Delta_1}{\Delta},\;\;\Delta=\left|
\begin{array}{cc}
\Psi^{(1;11)}_x &\Psi^{(1;11)}\cr
\Psi^{(2;11)}_x &\Psi^{(2;11)}
\end{array}
\right|=\\\nonumber
&&
\sum_{i,j=1}^2\Big((-1)^{i+1} K_1-(-1)^{j+1} K_2\Big)e^{\Big(\tilde\nu+(-1)^{i+1} K_1+(-1)^{j+1} K_2\Big)x}
\psi^{(11;i)}(y_1,y_2)\psi^{(21;j)}(y_1,y_2)
,\\\nonumber
&&
K_l= \frac{1}{2} \sqrt{\tilde \nu^2 +4 \tilde a \tilde \Lambda^{(l;1)} + 4\tilde \mu},\;\;y_1=z_1 -\tilde \Lambda^{(l;1)} t_1, \;\;y_2=z_2 -\tilde\Lambda^{(l;1)} t_2,\\\nonumber
&&
\Delta_1=\left|
\begin{array}{cc}
\Psi^{(1;11)}\tilde \Lambda^{(1;1)} &\Psi^{(1;11)}\cr
\Psi^{(2;11)}\tilde \Lambda^{(2;1)} &\Psi^{(2;11)}
\end{array}
\right|=\\\nonumber
&&
(\tilde \Lambda^{(1;1)}-\tilde \Lambda^{(2;1)})\sum_{i,j=1}^2e^{\Big(\tilde\nu+(-1)^{i+1} K_1+(-1)^{j+1} K_2\Big)x}\psi^{(11;i)}(y_1,y_2)\psi^{(21;j)}(y_1,y_2)
,\\\label{solution_b}
w^{(1;1)}&=&\frac{\Delta_2}{\Delta}
,\;\;
\Delta_2=\left|
\begin{array}{cc}
\Psi^{(1;11)}_x &\Psi^{(1;11)}\tilde \Lambda^{(1;1)} \cr
\Psi^{(2;11)}_x &\Psi^{(2;11)}\tilde \Lambda^{(2;1)} 
\end{array}
\right|.
\end{eqnarray}
Formulae (\ref{solution},\ref{solution_b})  have six arbitrary functions of two variables $\psi^{(l1;i)}(y_1,y_2)$, $\tilde \Lambda^{(l;1)}=E^{(l1)}(y_1,y_2)$,  $l,i=1,2$. It is evident that solution $u$ has no singularities if $\Delta\neq 0$, i.e., for instance, if 
\begin{eqnarray}
\label{non-sing}
K_1>K_2>0,\;\;\;\psi^{(11;2)}<0,\;\;\;
\psi^{(11;1)},\psi^{(21;1)},\psi^{(21;2)} >0,\;\;\forall \;y_1,y_2.
\end{eqnarray}
 Since $\tilde \Lambda^{(l;1)}$ is implicitly given by the eq.(\ref{nl_Lam_ex}), constructed function $u$ describes the break of the wave profile unless $\tilde \Lambda^{(l;1)}=const$. 
 
 Write the explicite form of the solution corresponding to 
 \begin{eqnarray}\label{K_0}
 &&
K_2=0\;\;\;\Rightarrow \;\;\;\tilde \Lambda^{(2;1)}=-\frac{\tilde \nu^2 +4 \tilde \mu}{4 \tilde a}=const, \\\nonumber
&&
 \psi^{(11;1)}(y_1,y_2)=\xi_1(y_1,y_2)>0,\;\; \psi^{(11;2)}(y_1,y_2)=-\xi_2(y_1,y_2)<0.
 \end{eqnarray}
 One has
\begin{eqnarray}\label{sol}
u=\frac{(\tilde \Lambda^{(1;1)}-\tilde \Lambda^{(2;1)})
\left(e^{K_1 x}\xi_1(y_1,y_2) -
e^{-K_1 x}\xi_2(y_1,y_2)\right)}
{K_1\left(e^{K_1 x}\xi_1(y_1,y_2) +
e^{-K_1 x}\xi_2(y_1,y_2)\right)},
\end{eqnarray}
 
 Let us consider the particular case   $\tilde \Lambda^{(l;1)}=const$, $l=1,2$. Then the solution to eq.(\ref{intr_v}) is given explicitely by the  eq.(\ref{solution}) with four arbitrary functions of two arguments $\psi^{(l1;i)}(y_1,y_2)$,  $l,i=1,2$. Consider solution given by formula (\ref{sol}) with $Im(K_1)=0$ and $Im(\xi_i)=0$, $i=1,2$. Function $u|_{z_i=const,t_i=const}$ is   kink. Behaviour of $u|_{x=const}$ as a function of $z_i$ and $t_i$ depends on the shapes of the arbitrary functions $\xi_i(y_1,y_2)$, $i=1,2$. In particular, $u_{x=const}$ may be  soliton.

\subsubsection{Simplest solutions to
eq.(\ref{h5_red_I})}
In accordance with eq.(\ref{Psi_sol_sc_red}), there is no arbitrary functions in the constructed solution space. The simplest nontrivial example of the solution to the eq.(\ref{intr_v}) is assotiated with  $N_0=n_0=1$.
As we have seen above, reduction (\ref{red_I}) requires $\tilde \Lambda^{(l;1)}=const$, $l=1,2$.  
Eq. (\ref{Psi_sol_sc_red})
reads
\begin{eqnarray}
\label{Psi_sol_sc_red_sol}
\Psi^{(l;11)}(\vec{x}) &=& \sum_{i=1}^2\psi^{(l1;i)}
e^{k^{(l1;i)} (x  -\tilde\Lambda^{(l;1)} y-(\tilde\Lambda^{(l;1)})^2 t)}
l=1,2,\;\;\psi^{(l1;i)}=const.
\end{eqnarray}
Eq.
(\ref{Psi1_ex}) remains correct. It yields
\begin{eqnarray}\label{solution2}
u&\equiv& v^{(1;1)}=\frac{\Delta_1}{\Delta},\;\;\\\nonumber
&&\Delta=\sum_{i,j=1}^2\Big((-1)^{i+1} K_1-(-1)^{j+1} K_2\Big)e^{\tilde\nu(q_1/2+q_2/2)+(-1)^{i+1} K_1 q_1+(-1)^{j+1} K_2 q_2}
\psi^{(11;i)}\psi^{(21;j)}
,\\\nonumber
&&
q_l(y,t)=x-\tilde \Lambda^{(l;1)} y - 
(\tilde \Lambda^{(l;1)} )^2 t ,\\\nonumber
&&
\Delta_1=
(\tilde \Lambda^{(1;1)}-\tilde \Lambda^{(2;1)})\sum_{i,j=1}^2e^{\tilde\nu(q_1/2+q_2/2)+(-1)^{i+1} K_1 q_1+(-1)^{j+1} K_2 q_2}
\psi^{(11;i)}\psi^{(21;j)}.
\end{eqnarray}
It is evident that solution $u$ has no singularities in the case (\ref{non-sing}). 

 Write the explicite form of the solution corresponding to conditions (\ref{K_0}) with constant $\xi_i$, $i=1,2$:
\begin{eqnarray}\label{sol2}
u=\frac{(\tilde \Lambda^{(1;1)}-\tilde \Lambda^{(2;1)})
\left(e^{K_1 q_1}\xi_1 -
e^{-K_1 q_1}\xi_2\right)}
{K_1\left(e^{K_1 q_1}\xi_1 +
e^{-K_1 q_1}\xi_2\right)},
\end{eqnarray}
If the parameter $K_1$ is real, then the function $u$ is  kink.

\section{Conclusions}
\label{Section:Conclusions}

The remarkable relations among $C$-, $Ch$-, $S_1$- and $S_2$-integrable nonlinear PDEs have been found in \cite{ZS_2} using a version of the dressing method.  
We represent a modification of this dressing algorithm  allowing one to involve a simple example  of the $S_3$-integrable scalar equation (i.e. eq.(\ref{intr_v}))  in the above list. In addition, a matrix version of this equation  is derived. We have found that (at least on the restricted manifold of solutions, as it is described in Sec.\ref{Section:Solutions})  eq.(\ref{intr_v}) originates  from the $GL(N^{S_2},\CC)$ SDYM  (which is  $S_2$-integrable equation) after the special differential reduction followed by the Frobenius type reduction imposed on the matrix field. In turn,  $GL(N^{S_2},\CC)$ SDYM may be derived from the (1+1)-dimensional hierarchy of $Ch$-integrable PDEs \cite{Z1,ZS_2}. Thus, 
the important result is that $S_3$-integrable equation (\ref{intr_v}) has been derived  from the (1+1)-dimensional hierarchy of nonlinear  $Ch$-integrable  PDEs by means of the following set of tree reductions imposed on the matrix fields:
\begin{eqnarray}\label{chain_3}
&&
{\mbox{FR}}\;\longrightarrow
{\mbox{DR($x$)}}\;\longrightarrow\;
{\mbox{FTR}}.
\end{eqnarray}
The first and the third reductions in eq.(\ref{chain_3}) introduce the new discrete variables into the  hierarchy of commuting  nonlinear PDEs. The second reduction  introduces the new independent variable $x$. After the third reduction, one must to  couple two   equations of the commuting hierarchy of chains in order to write the complete system of nonlinear PDEs.

Solution spaces to both  $S_2$-integrable PDEs with differential reduction and $S_3$-integrable  system are considered. It is shown that our algorithm allows one to construct a big manifold of solutions to eq.(\ref{intr_v}) and to its matrix generalization (\ref{NL1},\ref{NL3}) involving solutions with  break of the wave profile and soliton-like solutions. However, solution space to $S_3$-integrable PDEs available by our method is not full so that IVP may not be formally solved. The problem of improving this algorithm with the purpose to describe the full solution space is an important problem for further study.

 It is clear that the chain (\ref{chain_3}) may be prolonged:
\begin{eqnarray}\label{chain_3_2}
&&
{\mbox{FTR}}_1(l_1)\;\longrightarrow
{\mbox{DR}}_1(x_1)\;\longrightarrow
{\mbox{FTR}}_2(l_2)\longrightarrow
{\mbox{DR}}_2(x_2) \longrightarrow \cdots\\\nonumber
&&
\longrightarrow{\mbox{DR}}_{K-1}(x_{K-1}) \longrightarrow{\mbox{FTR}}_K (l_K).
\end{eqnarray}
Each   FTR$_i(l_i)$ introduces the new discrete parameter $l_i$,  while each DR$_i(x_i)$ introduces the new independent variable $x_i$ into the hierarchy of  chains of nonlinear PDEs. After the last reduction, one has to  couple two (or may be more)  chains from the  hierarchy  in order to write the complete system of nonlinear PDEs. Study of these PDEs and their possible relations with the classical integrable models is postponed for the future work.

Another evident modification of  our algorithm is generalization  of the differential reduction. Namely, instead of the second order linear PDE (\ref{Psi_xx}) one can use $K$th order linear PDE with arbitrary positive integer $K$:
\begin{eqnarray}\label{Psi_xx_g}
\partial_{x}^K\Psi^{(m)}=\sum_{j=0}^{K-1} \sum_{i=0}^{K-j-1}a^{(ij)} \partial_x^i\Psi^{(m)}  (\tilde\Lambda^{(m)})^j,\;\;m=1,\dots,2N_0.
\end{eqnarray}
Assotiated nonlinear PDEs will be considered elsewhere.

Author thanks Professor P.M.Santini for useful discussion. The work  was supported by the RFBR grants  
07-01-00446 and 06-01-92053 and by the grant NS-4887.2008.2.


\end{document}